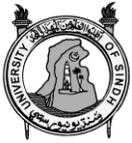
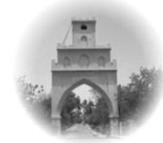

# A New Image Steganographic Technique using Pattern based Bits Shuffling and Magic LSB for Grayscale Images

K. MUHAMMAD++, J. AHMAD, H. FARMAN*, Z. JAN*

Department of Digital Contents, Digital Contents Research Institute, Sejong University, Seoul, South Korea.

**Abstract:** Image Steganography is a growing research area of information security where secret information is embedded in innocent-looking public communication. This paper proposes a novel crystographic technique for grayscale images in spatial domain. The secret data is encrypted and shuffled using pattern based bits shuffling algorithm (PBSA) and a secret key. The encrypted data is then embedded in the cover image using magic least significant bit (M-LSB) method. Experimentally, the proposed method is evaluated by qualitative and quantitative analysis which validates the effectiveness of the proposed method in contrast to several state-of-the-art methods.

## 1. INTRODUCTION

Steganography is the art and science of covert communication where secret data is hidden in innocent carriers, making it undetectable by human visual system (HVS). To increase the security of steganography, the secret information is encrypted using a cryptographic algorithm before message concealment. Steganography can be used for a number of useful applications including national and international governments' level security and secure banking and voting systems. In negative sense, it can be used by terrorists and criminals for their secret communication(Cheddad, Condell, Curran and Mc Kevitt 2010, Muhammad, Ahmad, Sajjad and Zubair 2015).

A number of different factors are considered while designing a new steganographic technique including payload, imperceptibility, and robustness. Payload is the amount of secret data that can be hidden in a particular carrier, measured in bits per pixel (bpp). Imperceptibility shows the chances of undetectability by HVS, which is measured using different image quality assessment metrics (IQAMs) like peak signal-to-noise ratio (PSNR), normalized cross correlation (NCC), and structural similarity index metric (SSIM)(Ahmad, Sajjad, Mehmood, Rho and Baik , Muhammad, Mehmood, Lee, Ji and Baik 2015). Robustness is the level of difficulty faced by attackers while extracting the hidden data from the stego image. The more security levels are introduced in a particular steganographic algorithm, the more robust is the steganography algorithm(Muhammad, Ahmad, Farman, Jan, Sajjad and Baik 2015).

The simplest spatial domain method of data hiding is LSB substitution, where the LSBs of the host image are replaced with the bits of secret data. Although the LSB method is quite simple and it results in good quality stego images but its detection is very easy. To decrease the asymmetric artefacts in stego images and make the detection of secret data difficult, LSB matching (LSB-M) is proposed(Luo, Huang and Huang 2010). LSB and LSB-M interpret the image pixels independently during data hiding which produces obvious artefacts in stego images. To solve this issue, (Mielikainen 2006) proposed LSB matching revisited (LSB-MR) which hides two secret bits at a time and takes into consideration the relationship between two neighboring pixels. LSB, LSB-M, and LSB-MR hide data directly in cover images without encryption. Furthermore, the image pixels are interpreted using raster scanning (starting from left-top and proceeding towards right-bottom). These two operations make the extraction of secret data easy for attackers. In addition, some of the existing techniques produce stego images of low quality, whose detection is relatively easy by HVS.

To address these limitations, this paper proposes a spatial domain based steganographic technique for grayscale images using M-LSB and PBSA. The main contributions of this paper are:

i) The proposed method introduces a new area of information security "Crystography" that is the combination of cryptography and steganography.

ii) The secret information is passed through PBSA based on a secret key before hiding it into the input image which increases the security of proposed method.

iii) The embedding algorithm used in this paper is a novel approach known as M-LSB that scatters the encrypted secret bits in the whole cover image to make the extraction for attacker more difficult.

The rest of the paper is organized as follows. Section 2 presents proposed methodology, followed by

---

++*Corresponding author,* K. MUHAMMAD: khanmuhammad@sju.ac.kr, khan.muhammad.icp@gmail.com
*Department of Computer Science, Islamia College Peshawar, Khyber Pakhtunkhwa, Pakistan.

experimental results in section 3. In section 4, the paper is concluded.

## 2. THE PROPOSED METHODOLOGY

In this section, we present the detailed novelty of the proposed method with suitable block diagram and easy to understand examples. The PBSA, embedding M-LSB method, and extraction algorithms are also presented in this section. The complete overview of the proposed steganographic framework is given in **(Fig. 1)**.

**Pattern based Bits Shuffling Algorithm**

The PBSA in the proposed model of steganography plays a vital role. It converts the secret data into bits representation and shuffles it based on a specific pattern and stego key. The use of PBSA makes the extraction of original hidden data difficult for attackers. The major steps of PBSA in pseudo code form are given in algorithm 1.

---

**Algorithm 1.** *Pattern based Bits Shuffling Algorithm*

**Input**: *Secret Information (M) and Secret Key (K)*
**Initialize:** *K ←key, M ←secret information, Msize← size(M),FinalBits← (length(M)*8), KeyBits← (length(K)*8), starting=1, j=8 and ending=8*

1. **for** *each character M(i) and K(i) in secret data M and stego key K* **do**
   a. *Convert M (i) into 8-bits & concatenate (FinalBits, 8-bits binary).*
   b. *Convert K(i) into 8-bits & concatenate (KeyBits, 8-bits binary).*
   **end for**
2. **for** *i ←1 to length (FinalBits)* **do**
      *temp ← FinalBits(i:i+7);*
      **for** *k ←1 to 4* **do**
         a. *tempVar ← temp(k);*
         b. *temp(k) ←FinalBits(j);*
         c. *FinalBits(j) ←tempVar;*
         d. *j←j-1;*
      **end for**
      *j←8;*
      *FinalBits(1:1,starting:ending)=temp(1:1,1:8);*
      *starting=ending+1;*
      *ending=starting+7;*
   **end for**

**Output**: *Cipher data*

---

**Embedding Algorithm**

The embedding algorithm embeds the encrypted secret data into the intensities of the host image. The embedding method is based on M-LSB substitution. M-LSB substitution plays an important role in scattering the secret data in the whole cover image which increases the security of the proposed method(Khan Muhammad 2015). The major steps of the proposed embedding algorithm in pseudo code form are given in algorithm 2.

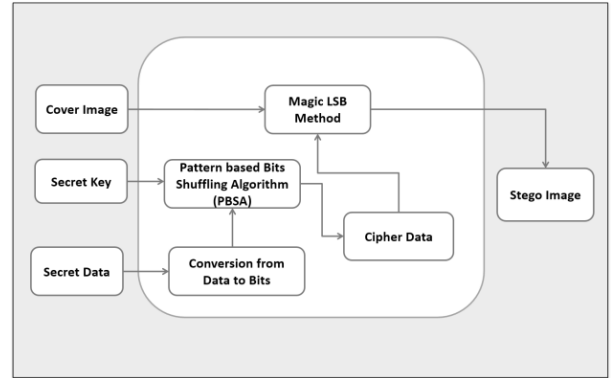

**Fig1: The proposed crystographic framework for data hiding**

---

**Algorithm 2.** *Embedding Algorithm*

**Input**: *Host Image ($I^H$), Secret information (M), Stego key (K)*

1. **Initialize** $I^H$ ←*Host image, M ←secret information, K← stego key*
2. *Apply Algorithm 1 on secret information M to get the cipher data.*
3. *Embed cipher data using Magic LSB as fellows:*
4. *Generate a magic matrix (MAGCM) with size equal to the size of host image $I^H$.*
5. **While** *counter <=size of message* **do**
   a. *Consider a pixel $I^H$ (i , j)*
   b. *Search for the index of a given message bit in MAGCM.*
   c. *Replace the LSB of the pixel at that particular index*
   d. *counter← counter +1;*
   **end**

**Output**: *Stego Image ($I^S$)*

---

To fully clarify the idea of magic LSB method, we present a simple example. Let's consider a host image $I^H$ consisting of nine (9) pixels {30, 46, 31, 65, 75, 22, 34, 98, and 59} and secret message bits $M= (01101011)_2$. For embedding this stream of bits, a magic matrix with size equal to the size of host image (3×3 in this case) is generated as shown below.

$$I^H: \begin{bmatrix} 30 & 46 & 31 \\ 65 & 75 & 22 \\ 35 & 98 & 59 \end{bmatrix} \quad MAGCM: \begin{bmatrix} 8 & 1 & 6 \\ 3 & 5 & 7 \\ 4 & 9 & 2 \end{bmatrix}$$

After hiding: $I^S: \begin{bmatrix} 31 & 46 & 30 \\ 65 & 75 & 23 \\ 34 & 78 & 59 \end{bmatrix}$

In the given example, the secret bits are embedded based on the indexes of MAGCM i.e. the first secret bit of M is hidden in pixel value (46), the second bit of M is embedded in 59, and the third bit of M is hidden in 65. Proceeding with the same way, the fourth bit is embedded in 35, fifth bit in 75, the sixth bit in 31, the seventh bit in 22, and the last 8$^{th}$ bit in pixel value 30. In the given example only 4 pixels changes which are shown in bold face. In case of image dimension 256×256, the MAGCM will be of dimension 256×256. At the receiver side, we have to extract the hidden bits according to the indexes of magic matrix and finally apply the reverse operations of PBSA.

**Extraction Algorithm**

The extraction algorithm extracts the embedded data from the stego image received from sender. The extracted bits are then decrypted using the reverse operations of PBSA. The major steps of extraction algorithm in pseudo code form are given in algorithm 3.

---

**Algorithm 3.** *Extraction Algorithm*

**Input**: *Stego Image ($I^S$), Stego key (K)*

1. **Initialize** $I^S$ ←*Stego image, K*← *stego key*
2. *Extract secret data using Magic LSB as fellows:*
3. *Generate a magic matrix (MAGCM) with size equal to the size of stego image $I^S$.*
4. **While** *size of message>=counter* **do**

    a. *Consider a pixel $I^S$ (i , j)*
    b. *Search for the 1$^{st}$ index in MAGCM (Start from 1$^{st}$ index and continue up to end of message size).*
    c. *Extract the LSB of the pixel at that particular index*
    d. *counter← counter +1;*
    **end**
5. *Apply the reverse operations of PBJA on extracted bits to get the original data.*

**Output**: *Secret Information (M)*

---

## 3. EXPERIMENTAL RESULT AND DISCUSSION

In this section, the proposed method is quantitatively and qualitatively evaluated and is compared with four methods including LSB method, Li et.al (Li, Yang and Zeng 2011) method, Peng at.al (Peng, Li and Yang 2012) method, and Yang et.al, (Gui, Li and Yang 2014) method. MATLAB version R2013a has been used for simulation. Some of the standard images used for experiments are shown in **(Fig. 2)**.

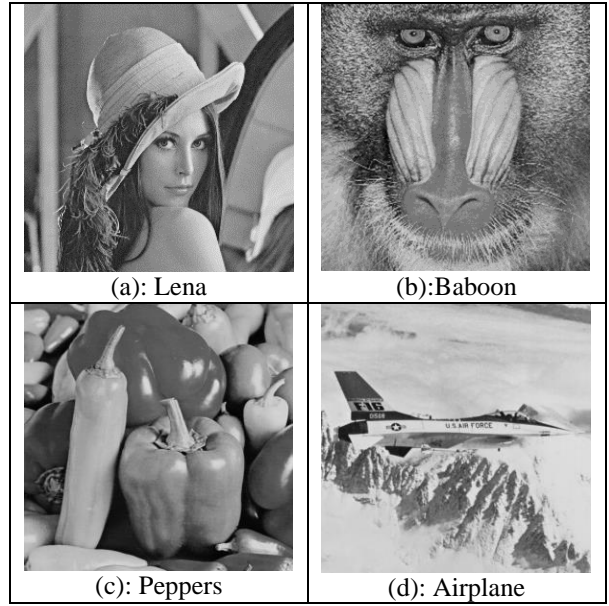

(a): Lena  (b): Baboon
(c): Peppers  (d): Airplane

**Fig2: Standard test images for experimental results**

**Quantitative Evaluation**

This section presents the quantitative evaluation of the proposed scheme and four other methods, based on the well-known IQAM PSNR. All the mentioned methods including the proposed scheme are tested with different embedding capacity (EC) based on PSNR in order to fully assess their performance. The PSNR is computed as follows(Muhammad, Ahmad, Farman and Zubair 2015, Muhammad, Sajjad, Mehmood, Rho and Baik 2015):

$$PSNR = 10\log_{10}\left(\frac{C^2_{max}}{MSE}\right) \quad (1)$$

$$MSE = \frac{1}{MN}\left(\sum_{x=1}^{M}\sum_{y=1}^{N}(S_{xy} - C_{xy})^2\right) (2)$$

Here, *M* and *N* show the image dimension, *S* shows the stego image, *C* shows the input cover image, and $C_{max}$ represents the maximum intensity in both images. $C_{max}$ is usually taken as 255 for 8-bit images(Mehmood, Sajjad, Ejaz and Baik 2015). The incurred quantitative results based on PSNR using different ECs in bits per pixel (bpp) are given in (**Table 1-3**).

Table 1 shows the quantitative results of proposed method and other methods based on PSNR. Each standard image is tested with EC=0.5 and its PSNR score is calculated as shown in Table 1. The average score of PSNR for each mentioned method is shown at the end of Table 1 in bold face. The average score of proposed scheme in terms of PSNR clearly dominates the existing state-of-the-art methods.

**Table1: Experimental results based on PSNR with embedding capacity=0.5**

| Image Name | LSB Method | Li (2011) | Peng (2012) | Yang (2014) | Proposed Method |
|---|---|---|---|---|---|
| Lena | 53.88 | 42.37 | 40.98 | 42.41 | 53.91 |
| Baboon | 53.88 | 31.42 | 30.31 | 32.01 | 53.89 |
| Barbara | 53.89 | 41.04 | 38.81 | 41.78 | 53.87 |
| Airplane | 53.91 | 45.97 | 44.17 | 46.03 | 53.88 |
| House | 53.91 | 44.92 | 43.05 | 45.25 | 53.92 |
| Lake | 53.90 | 36.62 | 31.18 | 37.15 | 53.90 |
| Boat | 53.90 | 37.82 | 36.90 | 37.81 | 53.88 |
| **Average** | **53.89** | **40.02** | **37.91** | **40.34** | **53.89** |

**Table2: Experimental results based on PSNR with embedding capacity=1**

| Image Name | LSB Method | Li (2011) | Peng (2012) | Yang (2014) | Proposed Method |
|---|---|---|---|---|---|
| Lena | 51.13 | 34.59 | 32.96 | 35.39 | 51.13 |
| Baboon | 51.14 | 23.81 | 22.35 | 24.32 | 51.15 |
| Barbara | 51.15 | 32.32 | 30.23 | 33.49 | 51.10 |
| Airplane | 51.12 | 37.76 | 35.96 | 38.51 | 51.15 |
| House | 51.12 | 34.89 | 33.45 | 36.90 | 51.15 |
| Lake | 51.13 | 29.78 | 29.14 | 31.20 | 51.14 |
| Boat | 51.16 | 30.68 | 29.32 | 31.20 | 51.13 |
| **Average** | **51.1403** | **31.97** | **30.48** | **33.00** | **51.1415** |

Table 2 shows the experimental results based on PSNR with EC=1 i.e. 1 bit per pixel for the proposed method and other existing mentioned methods. From Table 2, it is clear that Li et.al, method produces slightly better results than Peng et.al method. However, Yang et.al approach results in higher PSNR score as compared to Li and Peng approaches with EC=1. The proposed approach demonstrates better results than Li, Peng and Yang schemes and obtains slightly higher PSNR score than LSB method.

Table 3 shows the quantitative results of the proposed method and four existing methods with EC=1.5 using PSNR. The average score of PSNR is shown in bold face at the end of Table 3 for each mentioned scheme. Table 3 clearly shows that Peng et.al scheme results in lower score of PSNR for all standard test images. The PSNR score of Li et.al method is higher than Peng but smaller than Yang scheme. The average PSNR score of the proposed method is higher than existing four methods and hence results in better quality of stego images.

**Table3: Experimental results based on PSNR with embedding capacity=1.5**

| Image Name | LSB Method | Li (2011) | Peng (2012) | Yang (2014) | Proposed Method |
|---|---|---|---|---|---|
| Lena | 44.56 | 29.57 | 27.31 | 30.20 | 44.58 |
| Baboon | 44.58 | 23.81 | 22.35 | 24.32 | 44.59 |
| Barbara | 44.54 | 25.00 | 24.11 | 27.49 | 44.54 |
| Airplane | 44.58 | 32.08 | 29.96 | 32.90 | 44.60 |
| House | 44.53 | 28.76 | 26.09 | 30.54 | 44.56 |
| Lake | 44.56 | 29.78 | 29.14 | 31.20 | 44.58 |
| Boat | 44.59 | 25.12 | 23.44 | 25.80 | 44.61 |
| **Average** | **44.56** | **27.73** | **26.05** | **28.92** | **44.58** |

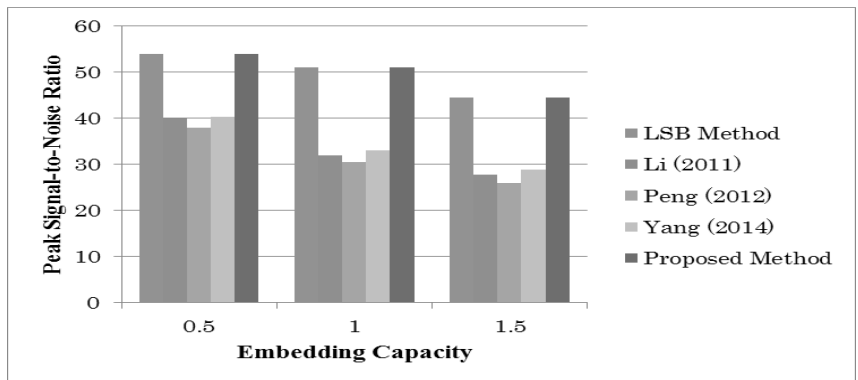

**Fig3: PSNR verses embedding capacity for all mentioned methods**

Fig. 3 shows the relationship between the EC and PSNR for the proposed scheme and other mentioned schemes. As PSNR computes the obvious distortion between the stego image and input cover image, therefore increasing the embedding capacity will decrease the value of PSNR(Mehmood, Sajjad and Baik 2014, Ahmad, Sajjad, Mehmood, Rho and Baik 2015). The PSNR score below 30dB shows low quality of stego images. So every steganographic algorithm should strive for PSNR higher than 30dB. An acceptable range of PSNR suggested by researchers of this area is PSNR>=40dB(Muhammad, Ahmad, Rehman, Jan and Qureshi 2015, Muhammad, Jan, Ahmad and Khan 2015). The proposed scheme satisfies this range despite of increasing the embedding capacity. However, Li, Peng and Yang schemes result in PSNR score smaller than 30dB which shows low quality of stego images. This indicates that the proposed scheme obtains high score of PSNR in all cases and hence validates its effectiveness.

**Qualitative Evaluation**

In this section, we evaluate the visual quality of stego images generated by the proposed scheme and other mentioned schemes based on HVS. The quality of a stego image is considered to be better if the detection of embedded data is extremely difficult using HVS. The visual quality assessment of stego images for the proposed scheme and other methods with different ECs is shown in (**Fig. 4**).

| Method Name | Cover Images: Lena and Baboon | Stego images with E.C=0.5 | Stego images with E.C=1 | Stego images with E.C=1.5 |
|---|---|---|---|---|
| Proposed Method | 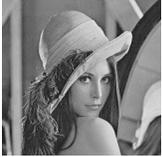 | 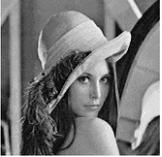 | 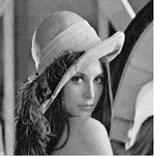 | 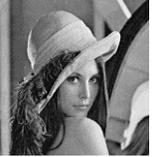 |
| LSB Method | 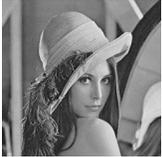 | 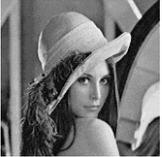 | 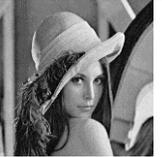 | 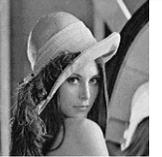 |
| Li et.al Method | 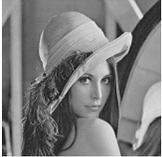 | 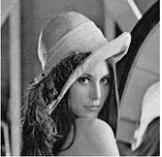 | 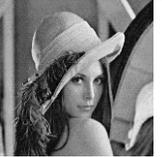 | 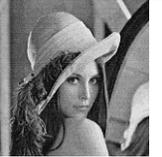 |
| Peng et.al Method | 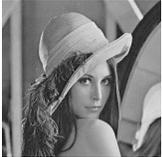 | 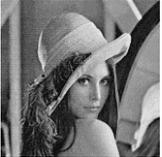 | 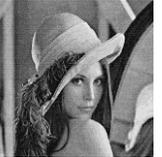 | 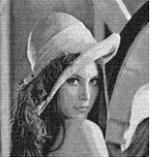 |
| Yang et.al Method | 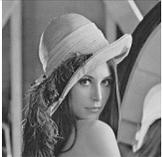 | 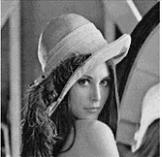 | 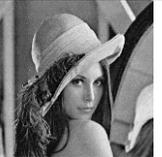 | 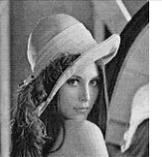 |

**Fig. 4: Visual quality assessment of stego images produced by various techniques. Column 1 describes all the mentioned methods. Colum 2 shows the test cover images of Lena. Column 3 represents the stego images of all mentioned methods with embedding capacity=0.5 bpp. Column 4 and column 5 present the stego images with payload= 1 and payload=1.5 bpp respectively.**

From Fig. 4, it can be validated that the stego images produced by Peng et.al method are fairly of low quality among other competing algorithms. The visual quality of stego images in case of Li's approach and

Yang's approach is almost same and better than Peng's approach. The stego images of LSB approach and proposed scheme are of high quality as compared to other three competing algorithms. The visual quality of stego images in the proposed scheme is slightly better than LSB scheme as validated by slightly larger PSNR score obtained by our proposed method in Table 1-3.

## 4. CONCLUSION

In this paper, an imperceptible image steganographic technique based on PBSA and M-LSB is proposed for grayscale images using spatial domain. The secret information is encrypted based on stego key and PBSA which is then embedded using M-LSB method in the cover image, scattering the secret data inside the whole image, hence making its extraction relatively more difficult for attackers. Experimentally, an acceptable average score of PSNR with 44.58dB is achieved, indicating the high quality of stego images, produced by our proposed technique. We conclude that the proposed technique is one of the candidates for secure communication over the Internet, maintaining the visual quality of stego images while ensuring the security of embedded data.

**REFERENCES:**

Ahmad, J., M. Sajjad, I. Mehmood, S. Rho and S. W. Baik "Saliency-weighted graphs for efficient visual content description and their applications in real-time image retrieval systems." Journal of Real-Time Image Processing: 1-17.

Ahmad, J., M. Sajjad, I. Mehmood, S. Rho and S. W. Baik (2015). "Describing Colors, Textures and Shapes for Content Based Image Retrieval-A Survey." Journal of Platform Technology 2 (4): 34-48.

Cheddad, A., J. Condell, K. Curran and P. Mc Kevitt (2010). "Digital image steganography: Survey and analysis of current methods." Signal processing 90(3): 727-752.

Gui, X., X. Li and B. Yang (2014). "A high capacity reversible data hiding scheme based on generalized prediction-error expansion and adaptive embedding." Signal Processing 98: 370-380.

Khan Muhammad, I. M., Muhammad Sajjad, Jamil Ahmad, Seong Joon Yoo, Dongil Han, Sung Wook Baik (2015). Secure Visual Content Labelling for Personalized Image Retrieval. The 11th International Conference on Multimedia Information Technology and Applications (MITA 2015) June 30-July2, 2015, Tashkent, Uzbekistan, Korea Multimedia Society.

Li, X., B. Yang and T. Zeng (2011). "Efficient reversible watermarking based on adaptive prediction-error expansion and pixel selection." Image Processing, IEEE Transactions on 20(12): 3524-3533.

Luo, W., F. Huang and J. Huang (2010). "Edge adaptive image steganography based on LSB matching revisited." Information Forensics and Security, IEEE Transactions on 5(2): 201-214.

Mehmood, I., M. Sajjad and S. W. Baik (2014). "Video summarization based tele-endoscopy: a service to efficiently manage visual data generated during wireless capsule endoscopy procedure." Journal of medical systems 38(9): 1-9.

Mehmood, I., M. Sajjad, W. Ejaz and S. W. Baik (2015). "Saliency-directed prioritization of visual data in wireless surveillance networks." Information Fusion 24: 16-30.

Mielikainen, J. (2006). "LSB matching revisited." Signal Processing Letters, IEEE 13(5): 285-287.

Muhammad, K., J. Ahmad, H. Farman, Z. Jan, M. Sajjad and S. W. Baik (2015). "A Secure Method for Color Image Steganography using Gray-Level Modification and Multi-level Encryption." KSII Transactions on Internet and Information Systems (TIIS) 9: 1938-1962.

Muhammad, K., J. Ahmad, H. Farman and M. Zubair (2015). "A novel image steganographic approach for hiding text in color images using HSI color model." Middle-East Journal of Scientific Research 22(5): 647-654.

Muhammad, K., J. Ahmad, N. U. Rehman, Z. Jan and R. J. Qureshi (2015). "A secure cyclic steganographic technique for color images using randomization." Technical Journal, University of Engineering and Technology Taxila 19(3): 57-64.

Muhammad, K., J. Ahmad, M. Sajjad and M. Zubair (2015). "Secure Image Steganography using Cryptography and Image Transposition." NED University Journal of Research 12(4): 81-91.

Muhammad, K., Z. Jan, J. Ahmad and Z. Khan (2015). "An Adaptive Secret Key-directed Cryptographic Scheme for Secure Transmission in Wireless Sensor Networks." Technical Journal, University of Engineering and Technology (UET) Taxila, Pakistan 20(3): 48-53.

Muhammad, K., I. Mehmood, M. Y. Lee, S. M. Ji and S. W. Baik (2015). "Ontology-based Secure Retrieval of Semantically Significant Visual Contents." Journal of Korean Institute of Next Generation Computing 11(3): 87-96.

Muhammad, K., M. Sajjad, I. Mehmood, S. Rho and S. W. Baik (2015). "A novel magic LSB substitution method (M-LSB-SM) using multi-level encryption and achromatic component of an image." Multimedia Tools and Applications: 1-27.

Peng, F., X. Li and B. Yang (2012). "Adaptive reversible data hiding scheme based on integer transform." Signal Processing 92(1): 54-62.